# Size dependence of the Zero Backward and Minimum Forward Scattering Conditions on Semiconductor Nanoparticles

Braulio García Cámara, José Francisco Algorri, Alexander Cuadrado[(1)], Virginia Urruchi,
José Manuel Sánchez-Pena, Ricardo Vergaz

Electronic Technology Department, Carlos III University of Madrid, E-28911 Leganés, Madrid.
Corresponding author phone: 0034916245964; fax: 0034916249430; e-mail: brgarcia@ing.uc3m.es.
Other emails: jalgorri@ing.uc3m.es, vurruchi@ing.uc3m.es, jmpena@ing.uc3m.es, rvergaz@ing.uc3m.es.

[(1)] Laser Processing Group, Instituto de Óptica, CSIC, C/Serrano 121, 28006 Madrid, Spain (email: a.cuadrado@csic.es)

**ABSTRACT**

The resonant modes observed in semiconductor nanoparticles and the coherence interaction between them, producing directional light scattering, may be very interesting for CMOS integrated all-optical devices. In these systems the control over the light scattering should be crucial, as well as the strength of this control. Fabrication parameters such as the size and shape of the nanoparticles and the optical properties of the environment can strongly affect to the emergence of these phenomena. In this work, we numerically explore the size dependence of the directional scattering conditions of semiconductor nanoparticles. Several semiconductor materials and a large size range have been considered to be a reference for further works. An interesting and unexpected linear behavior has been observed.

## 1. INTRODUCTION

Several technological advances in the field of photonics are currently focused on the development of all-optical devices for computing, storage or communications applications. They try to overcome the limitations of current electronic devices, like the communications bottleneck [1-3]. The miniaturization of computing systems is also one of the most important aspects in which photonic devices can highlight [4]. The achievement of all-optical devices for these applications, such as optical circuitry, requires the control of light scattering by these optical systems. In 1980's, Kerker and co-workers analytically obtained that light scattering of dipole-like particles can be suppressed in either the forward or the backward direction [5] by interaction of their electric and magnetic resonances. In addition, these phenomena were theoretically studied using complex arrangements of structures, known as metamaterials [6]. Recently, both zero-backward and minimum forward scattering were experimentally demonstrated using semiconductor nanoparticles [7-9]. In these works, the authors showed that semiconductor nanoparticles, like Silicon or Germanium, present both electric and magnetic resonances that may interact in the way described by Kerker. The control of light in the nanoscale as well as the CMOS compatibility of materials make these systems adequate to be integrated and/or to substitute current nanoelectronic devices. The emergence of these coherence effects requires a dominant dipolar behavior, such that only the first order Mie coefficients ($a_1$ and $b_1$) are not negligible. When this request is fulfilled, a zero backward scattering (ZB) can be achieved at the incident wavelength satisfying Eq. (1). On the contrary, a minimum forward scattering (MF) can be observed at wavelengths satisfying the expression of Eq. (2):

$$a_1 = b_1 \qquad (1)$$

$$a_1 = -b_1 \qquad (2)$$

Although Kerker and co-workers defined a zero-forward condition, current studies showed that only a minimum forward scattering can be achieved [7,10-11].

The resonant modes which allows the appearance of these phenomena behave as plasmon ones depending on the size, shape and the optical properties of the surrounding medium [11,12]. The shape is a critical parameter in this kind of systems, because the origin of these resonances is directly related to the geometry of the nanoparticles. While resonances are well-defined in spheres, they lose its sharp profile as the aspect ratio changes [13,14]. The dependence of both the spectral position of the resonances and their extinction efficiency on the surrounding medium have been already studied to use them as sensing parameters, following the works about plasmonic sensors [15]. The last important parameter, the particle size, has been also studied. A previous work [15] showed that while the electric resonances of semiconductor nanoparticles behaves as plasmon ones, they are red shifted as the particle size increases; the magnetic ones are more stable and their spectral position remains unalterable as the size of the particle increases. This asymmetric behavior should be reflected into the spectral position of the directional conditions as the particle size changes, presenting a size evolution different from that of resonances. In a recent work [16] an analytical relationship between the size and composition of the particle and the wavelength of the incident radiation was derived for the first Kerker's condition. In this work, we analyze, from a numerical point of view, the size dependence for both Kerker's directional conditions and for certain semiconductor materials presenting these interesting phenomena. The knowledge of this evolution can help further designs of all-optical devices based on a control of the scattered radiation.

2. METHODS

Mie theory [17] can be used to theoretically describe these phenomena in spherical nanoparticles. Assuming a dominant dipolar behavior, only the two first Mie coefficients, $a_1$ and $b_1$, are considered. The general expressions of these coefficients are given by:

$$a_n = \frac{m \Psi_n(mx) \cdot \Psi_n'(x) - \Psi_n(x) \cdot \Psi_n'(mx)}{m \Psi_n(mx) \cdot \xi_n'(x) - \xi_n(x) \cdot \Psi_n'(mx)} \quad (3)$$

$$b_n = \frac{\Psi_n(mx) \cdot \Psi_n'(x) - m \Psi_n(x) \cdot \Psi_n'(mx)}{\Psi_n(mx) \cdot \xi_n'(x) - m \xi_n(x) \cdot \Psi_n'(mx)} \quad (4)$$

Where $\Psi_n$ and $\xi_n$ are the n-order Ricatti-Bessel functions, $m$ is the relative refractive index of the nanoparticles with respect to the surrounding medium and $x$ is the size parameter, defined as $x = k \cdot r$, $k$ being the wavenumber of the incident beam and $r$ the radius of the particle.

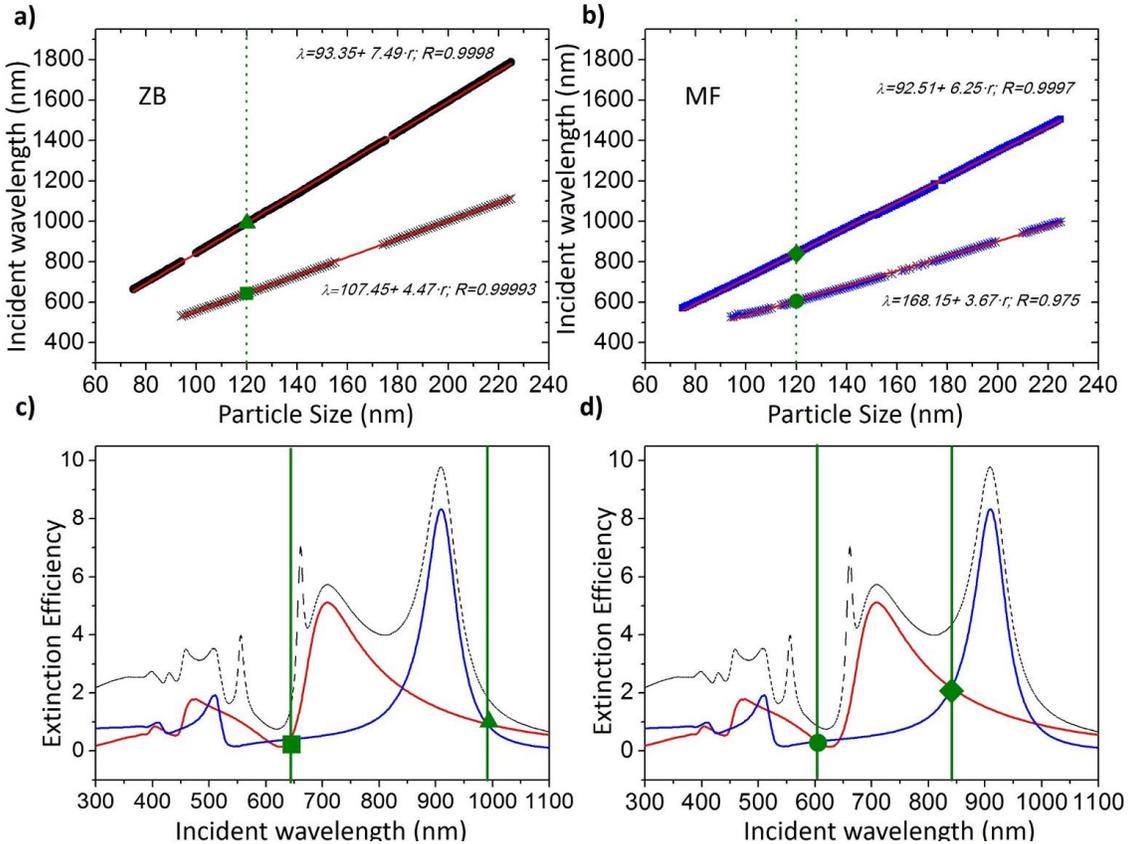

Fig. 1. Incident wavelength at which the zero-backward (a) or the minimum forward (b) conditions are satisfied in a Silicon spherical particle as a function of its radius. The two different curves in (a) and (b) correspond to different wavelength positions satisfying the directional conditions. The spectral dependence of the extinction efficiency of a Silicon sphere with r=120nm (black dotted line), as well as the contributions of first two dipolar modes, related to $a_1$ (red line) and $b_1$ (blue line) are plotted in (c) and (d). The two main wavelengths satisfying the zero-backward (c) or the minimum forward (d) are highlighted with green vertical lines: ZB at 643 nm and 992 nm and MF at 605nm and 841nm, respectively. The symbols used in the curves of (a) and (b) are superimposed in the corresponding cross they are referring. The linear fits of the curves of (a) and (b) are also included.

An iterative numerical method is used to find the Kerker's conditions (Eqs. (1) and (2)) as a function of the particle size, considering the particle embedded in air.

## 3. RESULTS AND DISCUSSION

By changing the radius of the spherical particle ($r$), it is possible to find the incident wavelengths satisfying Kerker's conditions (Eqs. (1) and (2)) as a function of $r$. Figures 1(a) and 1(c) show the size evolution of both the zero-backward (ZB, Eq. (1)) and the minimum forward (MF, Eq. (2)) conditions as a function of the particle size for a Silicon nanoparticle, respectively. The range of particle sizes was chosen in such a way that the directional behaviors appear in the visible and near-infrared ranges, in which photonic applications are currently developed.

While the electric dipolar resonance (related to $a_1$) follows a behavior with the nanoparticle size close to the Fröhlich theory [17], the magnetic dipolar resonance is almost motionless [15]. As the directional phenomena are consequence of a coherent interference between these contributions, their size evolution should be related to those of the dipolar modes. As can be seen in Fig. 1, this dependence has a fairly linear tendency. Linear fits of the curves are also included in Fig. 1. The slope of the curve in Fig. 1(a) is in accordance with the result shown in [16], where considering a constant refractive index of $n=3.5$ a relation of $x \cdot n = 2.7437$ (i.e. $\lambda/R = 8.015$, to compare with our 7.49) was obtained. In addition, we found two sets of values of $\lambda$ satisfying each directional condition, both following the linear behavior. These stacks are related to the main crosses of the dipolar contributions, both electric and magnetic. Figures 1(b) and 1(d) show the spectral dependence of the extinction efficiency (dotted black line) of a Silicon nanoparticle (r = 120 nm), and the dipolar contributions, related to the electric dipolar (red line) and the magnetic dipolar (blue line) modes. The wavelengths at which dipolar contributions cross each other are highlighted with vertical lines: while Fig. 1(b) remarks the two main wavelengths at which the zero-backward condition is satisfied, Fig. 1(d) does it the same with the minimum-forward wavelengths. There are more wavelengths fulfilling these conditions, however, the influence of higher-order modes avoid its view. In fact, low-wavelength set in both cases is influenced by the quadrupolar modes and the emergence of the directional phenomena is not warranted. In Figures 1(b) and 1(d), it can be seen that those low wavelengths (vertical lines on the left) are close to sharp peaks in the extinction efficiency (see black dotted line) with correspond to quadrupolar resonances [11,12,15]. This effect is also the main responsible of the lack of results at certain size ranges in Figs. 1(a) and 1(c).

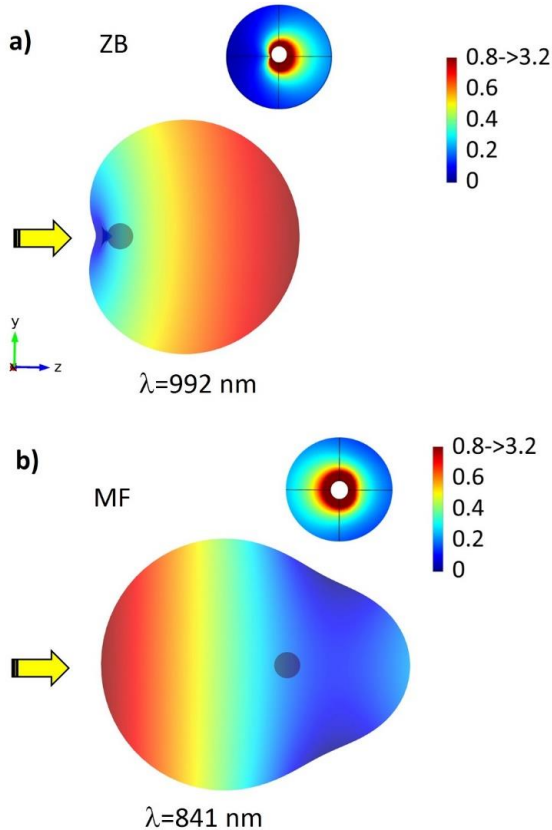

Fig. 2. Scattered electric field, in the far-field region, of a Silicon nanoparticle (r=120 nm) when it is illuminated by an incident beam with a wavelength satisfying (a) the zero-backward scattering condition and (b) the minimum-forward scattering condition. The corresponding wavelengths are obtained from previous curves. The scattered field in the near-field region has been also included in the insets. The incident beam is linearly polarized in the y direction and propagates in the z direction (remarked with a yellow arrow) with an amplitude of 1V/m. The position of the particle is included in the figures with a colored circle.

However, for large wavelengths, at which the dipolar behavior is dominant, the directional phenomena can be observed. For instance, we include in Fig. 2 FEM simulations (COMSOL Multiphysics ©) of a Silicon nanoparticle with a size and at two incident wavelengths fulfilling the directional conditions obtained above. In particular, Figure 2(a) shows the scattered field of a Silicon sphere with r = 120 nm illuminated by an incident beam with $\lambda$ = 992 nm, in such a way that a zero-backward scattering can be observed. Both the far-field and the near-field (inset) regions have been included. On the other hand, Figure 2(b) shows the scattered field of a Si sphere with r = 120 nm and a $\lambda$ = 841 nm in such a way that a minimum forward scattering can be observed. As it was mentioned previously, a zero forward scattering cannot be obtained, however a relevant reduction of the scattering in this direction and an enhancement of the backward scattering can be obtained, as it can be shown in Fig. 2(b). The directionality of the scattered field is more clearly observed in the far-field region than in the near field, due to the influence of the intense values of the electric field in the inside of the particle. However, a convenient choice of the color scale in the plot can also show the directionality in this region. Furthermore, the emergence of the zero-backward condition is more clearly compared with the minimum-forward one, in both spatial regions.

In a previous work [15] it was shown that these directional phenomena can be observed in nanoparticles made of other different semiconductor materials with high refractive index, like Germanium (Ge) or Gallium Arsenide (GaAs). The relative high refractive index of these materials [18] like that of Silicon, allows the view of these light phenomena. With the aim of generalization, the previous study has been also performed using these other semiconductor materials.

Figure 3 shows the wavelengths at which the directional conditions are satisfied in spherical nanoparticles made of different semiconductor materials. The size evolution of the directional phenomena in all of these materials also follows a linear behavior.

## 4. CONCLUSIONS

As a summary, we analytically analyzed the size dependence of the directional conditions, predicted by Kerker et al. [5] and recently observed in semiconductor nanoparticles. While plasmonic resonances in metallic nanoparticles follow the Fröhlich condition [17], the dependence of the dipolar resonances of semiconductor nanoparticles is slightly different; the electric one changes its spectral position with the size like plasmon resonances and the magnetic one remains at a certain wavelength or slightly shifts as the particle size changes. As the directional conditions are directly related to the spectral position of the dipolar resonances, the size dependence of them should be a joint effect. In particular, we have observed that both, the zero-backward condition and the minimum-forward conditions, linearly change with the particle size. In addition, we obtained two possible incident wavelengths satisfying the Kerker's conditions for each particle size. However, the presence of high order modes at low wavelengths avoids the emergence of the directional phenomena. This dependence has been observed for the main high-refractive semiconductors presenting electric and magnetic resonances in light scattering.

Currently, there are several research groups exploring the use of semiconductor nanoparticles to fabricate metadevices based on dielectric nanoparticles, like waveguides or nanoantennas. These results may help to further works to design new metadevices based on the directional effects.

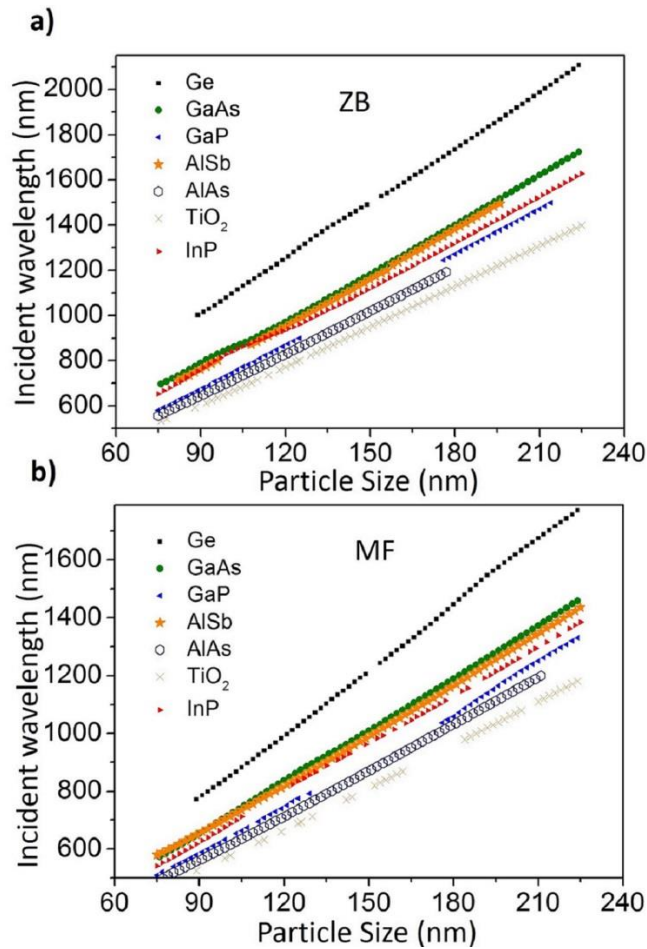

Fig. 3. Incident wavelength at which the zero-backward (a) or the minimum forward (b) conditions are satisfied in a spherical particle as a function of its radius and for several semiconductor materials presenting electric and magnetic resonances.

## REFERENCES

[1] B. García-Cámara, "Intra-/Inter-Chip Optical Communications" in *Communication Architectures for Systems on-Chip*, J. L. Ayala, Ed. Boca Ratón: CRC Press, 2011, pp. 249-332.
[2] M.A. Taubenblantt, "Optical Interconnects for High-Performance Computing," J. Light. Technol., vol. 30, no. 4, pp. 448-457, Feb. 2012.
[3] P. Chaisakul, D. Marris-Morini, J. Frigerio, D. Chrastina, M.S. Rouifed, S. Cechi, P. Crozat, G. Isella, and L. Vivien, "Integrated germanium optical interconnects on silicon substrates" Nature Photon., vol. 8, pp. 482-488, Mar. 2014.
[4] M. Naruse, *Nanophotonic Information Physics,* Berlin: Springer, 2014.
[5] M. Kerker, D.S. Wang and C.L. Giles, "Electromagnetic scattering by magnetic spheres", J. Opt. Soc. Am., vol. 73, no. 6, pp. 765-767, Jun. 1983.
[6] H. Caglayan, S.-H. Hong, B. Edwards, C.R. Kagan, N. Engheta, "Near-infrared metatronic nanocircuits by design", Phys. Rev. Lett., vol. 111, 073904, Aug. 2013.
[7] J. M. Geffrin, B. García-Cámara, R. Gómez-Medina, P. Albella, L.S. Froufe-Pérez, C. Eyraud, A. Litman, R. Vaillon, F. González, M. Nieto-Vesperinas, J. J. Sáenz and F. Moreno, "Magnetic and electric coherence in forward- and backward scattered electromagnetic waves by a single dielectric asubwavelength sphere" Nature Commum., vol. 3, 1171, Nov. 2012.
[8] Y. H. Fu, A.I. Kutnetsov, A.E. Miroshnichenko, Y.F. Yu and B. Luk'yanchuk, "Directional visible light scattering by silicon nanoparticles" Nat. Commum., vol. 4, 1527, Feb. 2013.
[9] S. Person, M. Jain, Z. Lapin, J.J. Sáenz, G. Wicks, and L. Novotny, "Demonstration of Zero Optical Backscattering from Single Nanoparticles," Nano Lett., vol. 13, no. 4, pp. 1806-1809, Mar. 2013.
[10] A. Alù, N. Engheta, "How does zero forward-scattering in magnetodielectric nanoparticles comply with the optical theorem?," J. Nanophoton., vol. 4, no.1, 041590, May 2010.
[11] R. Gómez-Medina, B. García-Cámara, I. Suarez-Lacalle, F. González, F. Moreno, M. Nieto-Vesperinas and J.J. Sáenz, "Electric and magnetic dipolar response of germanium nanospheres: interference effects, scattering anisotropy, and optical forces," J. Nanophoton., vol. 5, no.1, 053512, Jun. 2011.
[12] A. García-Etxarri, R. Gómez-Medina, L.S. Froufe-Pérez, C. López, L. Chantada, F. Scheffold, J. Aizpurua, M. Nieto-Vesperinas, and J.J. Sáenz, "Strong magnetic response of submicron Silicon particles in the infrared," Opt. Express, vol. 19, no. 6, pp. 4815-4826, Feb. 2011.
[13] K.A. Willets, and R.P Van Duyne, "Localized Surface Plasmon Resonance Spectroscopy and Sensing," Annu. Rev. Phys. Chem., vol. 58, pp. 267-297, May 2007.
[14] L. Staude, A.F. Miroshnichenko, M. Decker, N.T. Fofang, S. Liu, E. Gonzales, J. Dominguez, T.S. Luk, D.N. Neshev, I. Brener, and Y. Kivsnar, Y, "Tailoring Directional Scattering through Magnetic and Electric Resonances in Subwavelength Silicon Nanodisks" ACS Nano, vol. 7, no. 9, pp. 7824-7832, Aug. 2013.

[15] B. García-Cámara, R. Gómez-Medina, J.J. Sáenz, and B. Sepúlveda, "Sensing with magnetic dipolar resonances in semiconductor nanospheres," Opt. Express, vol. 21, no. 20, pp. 23007-23020, Oct. 2013.
[16] B. S. Luk`yanchuk, N. V. Voshchinnikov, R. Paniagua-Domínguez, A. I. Kuznetsov, "Optimum Forward Light Scattering by Spherical and Spheroidal Dielectric Nanoparticles with High Refractive Index". ACS Photonics. To be published, 2015.
[17] C.F. Bohren, and D.R. Huffman, *Absorption and Scattering of Light by Small Particles*, Weinheim: Wiley-VCH Verlag GmbH, 1983.
[18] E.D. Palik, *Handbook of Optical Constants of Solid*, San Diego, CA: Academic Press, 1985.

ACKNOWLEDGEMENTS

This work has been supported by Ministerio de Economía y Competitividad of Spain (grants no. TEC2013-47342-C2-2-R and no.TEC2013-50138-EXP) and the R&D Program SINFOTON S2013/MIT-2790 of the Comunidad de Madrid.